\documentclass{PoS}
\usepackage{amssymb}
\newcommand{\be}{\begin{eqnarray}}
\newcommand{\ee}{\end{eqnarray}}
\usepackage{subfigure}

\title{Generalized Parton Distributions of the Photon in Position Space}

\ShortTitle{Generalized parton distributions of the photon}

\author{\speaker{Asmita Mukherjee}\\
        Department of Physics, \\Indian Institute of Technology Bombay\\
         Powai, Mumbai 400076, India\\
        E-mail: \email{asmita@phy.iitb.ac.in}}

\author{Sreeraj Nair\\
        Department of Physics, \\Indian Institute of Technology Bombay\\
         Powai, Mumbai 400076, India}

\abstract{We report on a calculation of the generalized parton distributions of the photon when there is non-zero
momentum transfer both in the transverse and longitudinal directions. By taking Fourier transforms of the GPDs with respect to the
transverse and longitudinal momentum transfer, we obtain the parton distributions of the photon in
position space.}

\FullConference{ 10th International Symposium on Radiative Corrections (Applications of Quantum Field Theory to Phenomenology) - Radcor2011\\
September 26-30, 2011\\
Mamallapuram, India}

\begin{document}

\section{Introduction}
In recent years, experimental facilities have allowed access not only to inclusive processes like
deep inelastic scattering but also exclusive processes like the deeply virtual Compton
scattering (DVCS) $\gamma^{*}p \rightarrow
\gamma p$ or hard exclusive meson (M ) production (HEMP)$\gamma^{*}p \rightarrow M p$ .These processes are theoretically
described by the generalization of the parton distributions (pdfs)  to generalized parton
distributions (GPDs) 
\cite{rev1}\cite{rev2}\cite{rev3}\cite{rev4}.
The generalized parton distributions are expressed as  non-forward matrix elements of bilocal operators on the light-cone. 
These new distributions have been shown to be interesting theoretical tools for the study of hadrons and they connect 
through sum rules, to the hadronic form factors. Also their forward limit connects to the usual
pdfs.
Much of the interest in these quantities has been triggered by their potential to help unravel the spin structure 
of the nucleon, as they contain information not only on the helicity carried
by partons, but also on their orbital angular momentum.
The off-forward nature of the GPDs requires an additional variable for the description of the initial and final
proton states,
called the skewness variable, $\zeta$, which reflects the longitudinal momentum
transfer.
Hence the generalized parton distributions, at a given scale,  are functions of three
variables namely they depend on $x$, 
the longitudinal momentum fraction of the struck parton, $\zeta$ and the square of the momentum transfer in the process ($-t$).
\newline
\hspace*{0.5cm}The photon is one of the fundamental gauge bosons of the Standard Model without self couplings and without 
intrinsic structure. But at high energies when the virtuality of the probe is large the interactions are dominated by 
quantum fluctuations of photon into fermion-antifermion pairs. This is called photon structure.
The photon structure function is understood fairly accurately and agrees well with experimental results \cite{buras}.
For the GPDs of the photon, deeply virtual Compton scattering on a photon target
$\gamma^{*}\gamma \rightarrow \gamma \gamma$ was considered in \cite{pire} in the kinematic region of large $Q^2$
but small squared momentum transfer ( $-t < <  Q^2$).
The momentum transfer was taken purely to be in longitudinal direction ($\Delta_\perp = 0 $) in \cite{pire} and the
photon GPDs were related with the Fourier transform of the matrix elements of light-front bilocal 
currents between initial and final photon states.
The case when the momentum transfer is purely in the transverse direction ($\Delta_\perp \ne 0, \zeta = 0 $)
 was studied in \cite{us1}. We consider here the case when there is non-zero momentum transfer both in the transverse
 and longitudinal directions ($\Delta_\perp \ne 0, \zeta \ne 0 $).
 As the GPDs involve a momentum transfer (off-forwardness), they do not have
probabilistic interpretation, unlike ordinary parton distributions (pdfs). However, it has been
shown that when the momentum transfer is purely in the transverse direction ($\zeta = 0$), if one performs
a Fourier Transform (FT) with respect to the transverse momentum transfer $\Delta_\perp$, one gets the
so-called impact parameter dependent parton distributions $q(x,b^\perp)$ (ipdpdfs)
 \cite{burkardt}\cite{burkardt1}\cite{burkardt2}.
Ipdpdfs of the photon tell us how the quarks of a given longitudinal momentum fraction $x$ are distributed in the 
transverse position  or impact parameter space at a distance $b^\perp$ from the center of the photon.
These obey certain positivity conditions and unlike the GPDs, have probabilistic interpretation.
We consider here the ipdpdfs of the photon when $\zeta \ne 0$. Unlike the case when $\zeta = 0$, photon
GPDs do not have a probabilistic interpretation when $\zeta \ne 0$,
however they now describe the parton distributions when the initial photon is displaced from the final photon 
in the transverse impact parameter space. The same description holds for proton GPDs as shown in \cite{proton}.
We also consider the parton distribution of the photon in the longitudinal position space by introducing a 
longitudinal impact
parameter $\sigma$ conjugate to the skewness $\zeta$. For an electron dressed with a photon in QED, it was found that 
the DVCS amplitude show  a diffraction-like pattern in longitudinal position space 
\cite{hadron_optics}\cite{hadron_optics1}.
Similar diffraction patterns were observed for proton GPDs in the longitudinal position space in 
\cite{model}\cite{model1} which were found to be model dependent.
We consider here the photon GPDs in both the transverse as well as longitudinal position space when 
the skewness is non-zero.

\section{GPDs of the photon for non-zero skewness}

For a real photon target state, we can define the photon GPDs as the following non-forward matrix elements
\cite{pire}:


\begin{figure}
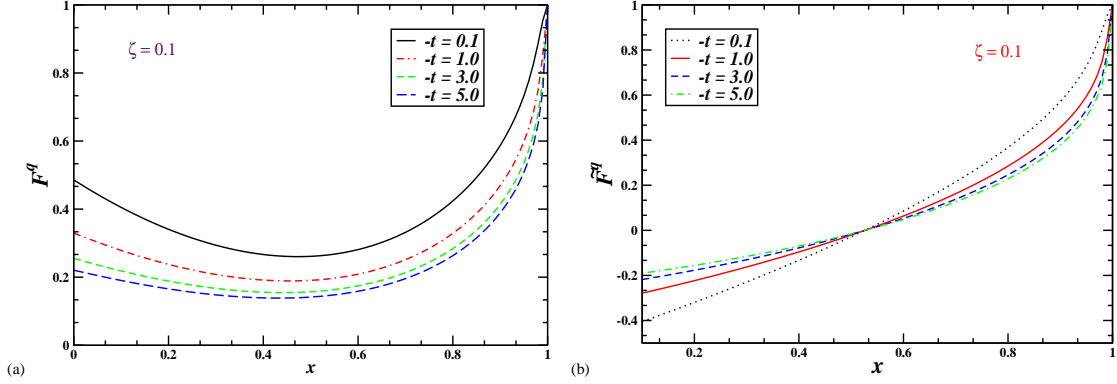

\centering
\mbox{\subfigure{\tiny{(a)}\includegraphics[width=7cm,height=5cm,clip]{fig1a.eps} 
\quad
\subfigure{\tiny{(b)}\includegraphics[width=7cm,height=5cm,clip]{fig1b.eps}}}}  
\caption{(Color online)  (a) Plot of unpolarized
GPD $F^{q}$ vs $x$ for fixed values of $-t$ in $GeV^{2}$ and at $\zeta=0.1$
and (b) polarized GPD  $\tilde{F^{q}}$ vs $x$ for fixed values of $-t$ in
$GeV^{2}$ and at $\zeta=0.1$, $\Lambda = 12 \mathrm{GeV}$. }
\label{fig1}

\end{figure}


\be
F^q=\int {dy^-\over 8 \pi} e^{-i P^+ y^-\over 2} \langle \gamma(P') \mid
{\bar{\psi}} (0) \gamma^+ \psi(y^-) \mid \gamma (P)\rangle ;
\nonumber\\ 
\tilde F^q=\int {dy^-\over 8 \pi} e^{-i P^+ y^-\over 2} \langle \gamma(P') \mid
{\bar{\psi}} (0) \gamma^+ \gamma^5 \psi(y^-) \mid \gamma (P)\rangle .
\ee

$F^q $ and $\tilde F^q$ give the GPDs for unpolarized and polarized photon respectively.
We choose the light-front gauge $A^{+} = 0$. We consider the region  $1>x>\zeta$ and $-1<x<\zeta-1$ 
which corresponds to the diagonal contributions coming from the  particle number conserving overlap
of the light-front wave functions \cite{us2}. 

We work in the standard light-cone co-ordinates $x^\pm = x^0 \pm x^3$.
We choose our frame such that:
\begin{eqnarray}
P&=&
\left(\ P^+\ ,\ {0^\perp}\ ,\ 0\ \right)\ ,
\label{a1}\\
P'&=&
\left( (1-\zeta)P^+\ ,\ -{\Delta^\perp}\ ,\ {{\Delta}^{\perp 2} \over (1-\zeta)P^+}\right)\ ,
\end{eqnarray}
The four-momentum transfer from the target is
\begin{eqnarray}
\label{delta}   
\Delta&=&P-P'\ =\
\left( \zeta P^+\ ,\ {\Delta^\perp}\ ,\
{t+{\Delta^\perp}^2 \over \zeta P^+}\right)\ ,
\end{eqnarray}
where $t = \Delta^2$. In addition, overall energy-momentum
conservation requires $\Delta^- = P^- - P'^-$, which connects 
${\Delta^\perp}^2$, $\zeta$, and $t$ according to
\begin{equation}
 (1-\zeta) t  = -{\Delta^\perp}^2 .
 \label{tzeta}
\end{equation}

The quark contribution comes from $1>x>\zeta$ region and the anti-quark contribution comes from $-1<x<\zeta-1$ region.
The photon GPDs are calculated in terms of the light-front wave functions of the target photon, which can be evaluated
in perturbation theory \cite{us2}.

\section{Photon GPDs in impact parameter space}

Parton distribution of photon in the transverse impact parameter space for non-zero skewness is defined as :

\be
q(x,\zeta,b)={1\over 4 \pi^2} \int d^2 \Delta^\perp e^{-i \Delta^\perp 
\cdot b^\perp} F^q (x,\zeta,t)\nonumber\\
\tilde q (x,\zeta,b)={1\over 4 \pi^2} \int d^2 \Delta^\perp  e^{-i \Delta^\perp
\cdot b^\perp} \tilde F^q (x,\zeta, t).
\ee
As described in the introduction the parton distributions in impact parameter space describe the probability of 
finding a parton of definite momentum fraction $x$ at a distance $ b^\perp$ from the center of the photon.
Here $b= \mid b^\perp \mid $ is the transverse  impact parameter which is a
measure of  the transverse distance between the struck quark and the center of momentum of the photon.
Unlike the case when skewness is zero, for non-zero skewness the transverse location of the photon is not the same 
before and after
the scattering. This shift is independent of $x$ but depends on the skewness $\zeta$ and $ b^\perp$ as a result 
the information related to this transverse shift
does not vanish  when the GPDs are integrated over $x$. The impact parameter dependent pdfs would have been a 
delta function for a single quark.
Thus the smearing in the $b_\perp$ space reveals the partonic content of the photon.


\begin{figure}
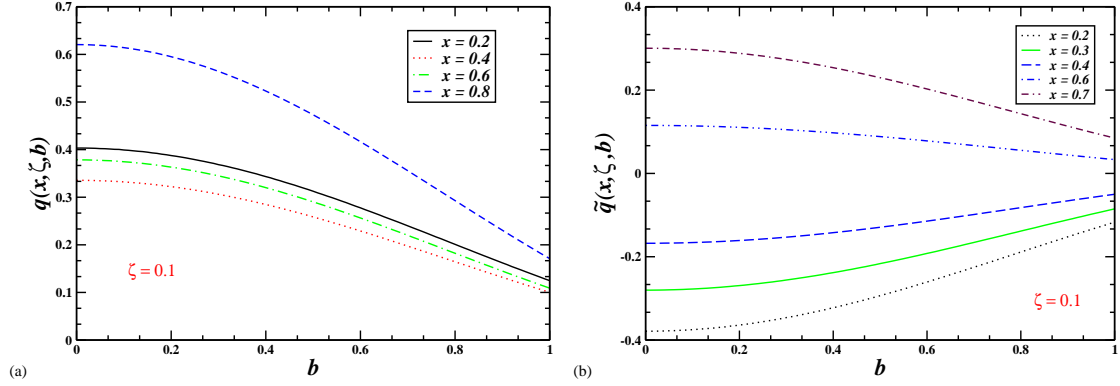

\centering
\mbox{\subfigure{\tiny{(a)}\includegraphics[width=7cm,height=5cm,clip]{fig2a.eps}
\quad
\subfigure{\tiny{(b)}\includegraphics[width=7cm,height=5cm,clip]{fig2b.eps} }}}

\caption{(Color online) (a) Plot of $q(x,\zeta,b)$ vs $b$  and (b) $\tilde q(x,\zeta,b)$ vs $b$ for fixed
 values of $x$ and at $\zeta = 0.1$,where we have taken 
$\Lambda$ = 12 $\mathrm{GeV}$ and $\Delta_{max}$= 3~GeV where $\Delta_{max}$ is the upper limit in
the $\Delta$ integration. The distributions are given in $\mathrm{GeV}^2$ and 
$b$ is in ${\mathrm{GeV}}^{-1}$. }
\label{fig2}

\end{figure}


We introduce a coordinate $b$ conjugate to the momentum transfer $\Delta$ such that $ b .\Delta =
\frac{1}{2}b^{+} \Delta^{-} + \frac{1}{2}b^{-} \Delta^{+} -  b_{\perp} \Delta_{\perp}$.
Also $\frac{1}{2}b^{-} \Delta^{+} = \frac{1}{2}b^{-} P^{+}\zeta = \sigma \zeta$ where we have
defined the boost invariant variable $\sigma$ which is an 'impact parameter' in the longitudinal
position space. The Fourier transform of the GPDs with respect to $\zeta$ allows
one to determine the longitudinal structure of the target hadron in terms of the variable $\sigma$.

The photon GPDs in longitudinal position space is given by:
\be
q(x,\sigma, t)= \frac{1}{2\pi}\int_0^{\zeta_{max}} d\zeta e^{i \zeta \sigma} 
F^q (x,\zeta,t)\nonumber\\   
\tilde q (x,\sigma, t)=\frac{1}{2\pi}\int_0^{\zeta_{\max}} d\zeta e^{i \zeta \sigma} \tilde F^q (x,\zeta,t)
\ee
The upper limit of $\zeta$ integration $\zeta_{max}$ comes out as $x$ in the region $\zeta<x<1$.
It was shown that DVCS amplitude shows a diffraction pattern in longitudinal impact parameter space in 
\cite{hadron_optics}\cite{hadron_optics1}. Similar diffraction 
patterns were observed for the proton GPDs in the longitudinal position space in \cite{proton}.
Photon GPDs also show a diffraction like pattern in the longitudinal position space with distinct 
features as compared to the proton GPDs.


\begin{figure}
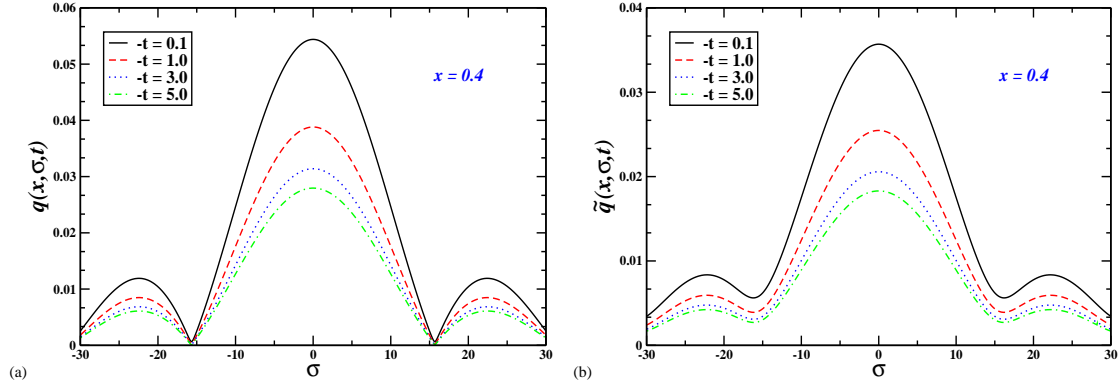

\centering
\mbox{\subfigure{\tiny{(a)}\includegraphics[width=7cm,height=5cm,clip]{fig3a.eps}
\quad
\subfigure{\tiny{(b)}\includegraphics[width=7cm,height=5cm,clip]{fig3b.eps} }}}

\caption{\label{fig3}(Color online) (a) Plot of $q(x,\sigma,t)$ vs $\sigma$  for
 a fixed value of $x=0.4$ and different values of $-t$ in $GeV^{2}$ and
 (b) $\tilde q(x,\sigma,t )$ vs $\sigma$ for a fixed
value of $x=0.4$ and different values of $-t$ in $GeV^{2}$ we have taken $\Lambda$ = 12 $\mathrm{GeV}$.}

\end{figure}


\section{Results}

The nature of the photon GPDs for non-zero skewness with respect to $x$ is same as that observed for the case when
skewness is zero \cite{us1}. This is true for both the polarized and
unpolarized photon GPDs.
Both $F^q$ and $\tilde F^q$ become independent of $t$ as $x\rightarrow 1$ as seen in Figs. 1(a) and (b)
because in this limit all the momentum is carried by the quark in the photon.
We took the upper limit in the momentum integration $\Lambda = 12 \mathrm{GeV}$ and the quark mass 
to be $m = 0.0033 \mathrm{GeV}$. The Fourier transform of $F^q$ and $\tilde
F^q$ with respect to $\Delta^\perp$ for non-zero $\zeta$ as a function of $b$ and a fixed value of $x$
and for different values of $-t$ is shown in Figs. 2 (a) and (b).
The GPDs in transverse impact parameter space for non-zero $\zeta$ probe partons inside the target photon
when the initial photon is displaced from the final photon in the transverse impact parameter space.
The upper limit in the $\Delta_\perp$ integration is taken to be $\Delta_{max} = 3
 \mathrm{GeV}$.
The Fourier transform of $F^q$ and $\tilde F^q$ with respect to $\zeta$ as a
function of the boost invariant 
variable $\sigma$ at fixed value of $x$
and for different values of $-t$ is shown in Figs. 3(a) and (b).
The unpolarized photon GPD shows diffraction pattern in the longitudinal 
position space with a central maxima and 
several secondary maxima separated by well-defined minima as in Fig. 3(a). However no such prominent diffraction 
pattern is observed for the polarized GPDs as seen in Fig. 3(b).
An analogy with the diffraction pattern in optics was given in \cite{hadron_optics}\cite{hadron_optics1}.
The finite range of $\zeta$ integration acts as a slit of finite width necessary to produce the diffraction pattern.
The height of the maxima of both $q(x, \sigma,t) $ and $\tilde q(x,\sigma, t)$ decrease with increasing $-t$.

\section{Conclusions}

We calculated the photon GPDs, both polarized and unpolarized in the general case when the skewness parameter 
was non-zero and 
when the momentum transferred was both in the longitudinal as well as transverse directions.
We considered the distribution of the partons in the impact parameter space both in the longitudinal and 
transverse direction.
We calculated the photon GPDs for the case when the helicity of the initial and the final photon states 
are the same or in other words the non-helicity-flip photon GPDs.
The photon GPDs in the transverse impact parameter space for non-zero skewness shows the same features as observed
for the photon GPDs when skewness is zero \cite{us1}.
The photon GPDs in the longitudinal impact parameter space show interesting diffraction pattern analogous to those 
observed in optics. The diffraction pattern appears because of the finiteness of the $\zeta$ integration and 
it depends on $x$ and $-t$.

\section{Acknowledgments}

This work is supported by BRNS grant Sanction No. 2007/37/60/BRNS/2913 dated 31.3.08,
Govt. of India. AM thanks the organizers of RADCOR for the invitation.

\end{document}